# Turbulent Schmidt Number Measurements Over Three-Dimensional Cubic Arrays

Annalisa Di Bernardino[1] · Paolo Monti[2]* · Giovanni Leuzzi[2] · Giorgio Querzoli[3]

[1] Dipartimento di Fisica, Università di Roma "La Sapienza", Piazzale Aldo Moro 2, 00185, Rome, Italy
[2] DICEA, Università di Roma "La Sapienza", Via Eudossiana 18, 00184, Rome, Italy
[3] DICAAR, Università degli Studi di Cagliari, Via Marengo 2, 09123, Cagliari, Italy
* e-mail: paolo.monti@uniroma1.it

**Abstract** We present turbulent Schmidt number ($Sc_t$) estimations above three-dimensional urban canopies, where $Sc_t$ is a property of the flow defined as the ratio of the eddy diffusivity of momentum ($K_M$) to the eddy diffusivity of mass ($D_t$). Despite the fact that $Sc_t$ modelling is of great interest, inter alia, for pollutant dispersion simulations conducted via computational fluid dynamics, no universal value is known. Simultaneous measurements of fluid velocity and mass concentration are carried out in a water channel for three staggered arrays of cubical obstacles corresponding to isolated flow, wake-interference, and skimming-flow regimes. A passive tracer is released from a continuous point source located at a height $z = 1.67H$, where $H$ is the obstacle height. The results show an increase of $Sc_t$ with height above the canopy for all three arrays, with values at $z = 2H$ ($Sc_t \approx 0.6$) about double compared to that at $z = H$. The observed $Sc_t$ agrees well with that modelled by using a simple formulation for $Sc_t$ based on expressions for $K_M$ and $D_t$ published in previous studies. Comparisons with other $Sc_t$ models found in the literature are also presented and discussed.

## 1 Introduction

Urban canopies are heterogeneous in nature and have geometrical and thermal properties that play a fundamental role in air quality, human comfort, and energy consumption (Fernando 2010; Barlow 2014; Buccolieri et al. 2015; Salvati et al. 2019). Owing to the presence of street canyons, squares, intersections, and buildings of varying heights, estimation of air pollutant concentration in the urban environment has proved to be a difficult task to achieve (e.g., Ben Salem et al. 2015). That is particularly true if the analysis is intended to give parametric laws that are easy to use and capable of predicting pollutant concentration fields (Di Sabatino et al. 2008; Huq and Franzese 2013; Goulart et al. 2018). For this reason, the real urban environment is usually simplified by adopting arrays of obstacles with archetypal arrangements in both numerical and laboratory studies (Zajic et al. 2011).

In the last few decades, results from field campaigns, laboratory experiments, and numerical simulations have made it possible to assess how the presence of obstacles alters both the airflow and the concentration field (Querzoli et al. 2017). However, only in recent years have there been studies investigating the mechanisms governing the turbulent exchanges of mass and momentum (see e.g., Baik and Kim 2002; Barlow et al. 2004; Kanda 2006; Salizzoni et al. 2009; Takimoto et al. 2011; Neophytou et al. 2014; Nosek et al. 2016; Tomas et al. 2017).

Pollutant dispersion has been widely investigated through computational fluid dynamics (CFD, see e.g., the recent review by Lateb et al. 2016). Since these studies are generally performed using Reynolds-averaged Navier–Stokes (RANS) models or large-eddy simulation (LES) models (Badas et al. 2017), the question arises regarding how to deal with the turbulent mass and momentum fluxes remaining in the governing equations. These unknowns are currently modelled by appealing to first-order closure, which establishes a direct dependence of the generic turbulent flux on the gradient of the averaged transported quantity (momentum or mass) and on an exchange coefficient, e.g., the eddy diffusivity of momentum ($K_M$) or mass ($D_t$). In current practice, $K_M$ is assumed proportional to the turbulence kinetic energy ($q$) and to its rate of dissipation ($\varepsilon$), viz. $K_M = C_\mu q^2/\varepsilon$, where $C_\mu$ is a constant of proportionality, while $D_t = K_M(Sc_t)^{-1}$, where $Sc_t$ is the turbulent Schmidt number. The choice of $Sc_t$ is not straightforward (Di Sabatino et al. 2007; Tominaga and Stathopoulos 2011) and the selected value has a considerable influence on the computed concentrations (Tominaga and Stathopoulos 2007). It must be set prior to the simulation and it is generally assumed to fall in the 0.2–1.3 range (Gualtieri et al. 2017). The Schmidt number is sometimes determined after a series of CFD validations conducted at different $Sc_t$ values against experimental data (Blocken et al. 2008). Tominaga and Stathopoulos (2012) evaluated $Sc_t$ to be used in a RANS model referring to the results from a LES model, and found significant $Sc_t$ inhomogeneities within the canopy, with values varying in the range 0.2–2. Based on comparisons between CFD and experimental data, Riddle et al. (2004), Di Sabatino et al. (2007) and Nakiboglu et al. (2009) found a good agreement between measured and simulated concentrations for $Sc_t = 0.4$. Furthermore, Ebrahimi and Jahangirian (2013) found $Sc_t = 0.7$ to be the best value that led to the most accurate results using a RANS model. Recently, Gorlé et al. (2010) and Longo et al. (2019) proposed formulations for $Sc_t$ whose coefficients depend on turbulence characteristics and position.

Since experimental estimates of $Sc_t$ require information on $D_t$, it is necessary to conduct measurements aiming to directly quantify turbulent fluxes of scalars. Nevertheless, owing to intrinsic problems related to the employment of intrusive sensors, measuring simultaneously fluid velocity and scalar concentration at several points of the domain of interest with the necessary spatial and temporal resolution is not straightforward (Carpentieri et al. 2012, and 2018; Nosek et al. 2017). Direct measurements of $Sc_t$ have been performed in a wind tunnel simulating an atmospheric boundary layer above flat terrain by Koeltzsch (2000), who found a clear dependence of $Sc_t$ on height, i.e., an increase from the surface up to $Sc_t \approx 1$ at about 1/3 of the boundary-layer depth followed by a decrease in the overlaying region. As argued by Flesh et al. (2002), variations of $Sc_t$ with height could partially explain the wide variability found in reported $Sc_t$ values.

In recent years, the use of image analysis techniques has permitted the determination of turbulent scalar fluxes in water channels or in wind tunnels along planes instead of single points (see e.g., Vinçont et al. 2000; Dezső-Weidinger et al. 2003; Monti et al. 2007; Tomas et al. 2017; Di Bernardino et al. 2018). With the aim of evaluating $Sc_t$ and understanding how it is related to turbulence structures, the present study follows that methodology and provides information on the turbulent exchanges of scalars associated with stationary emission of a passive non-buoyant and non-reactive tracer from a point source located above a staggered array of cubes.

In the present work, $Sc_t$ is determined based on the gradient transport theory and for this reason the analysis focusses on the region above the canopy. Within the canopy the turbulent fluxes can possibly be downgradient (e.g., Dezső-Weidinger et al. 2003) and the gradient transport theory would, therefore, be inapplicable.

The paper is organized as follows. Section 2 describes the modelling set-up and the experimental study undertaken, while Sect. 3 presents the results obtained for three different cubical arrays. In Sect. 3 we also present a simple model for $Sc_t$ based on known expressions for $K_M$ and $D_t$ and parameters derived from the present experiments. Section 4 is devoted to the comparison of the $Sc_t$ values obtained using our model with those calculated using the Gorlé et al. (2010) and Longo et al. (2019) models. Section 5 concludes with a summary of the main findings.

## 2 Experiments

Here, only the main features are recalled since details of the experimental set-up, acquisition techniques, and data processing have been presented elsewhere (Di Bernardino et al. 2015, 2017, 2018, and 2019). The experiments were conducted in the recirculating, constant-head, water channel of the Hydraulic Laboratory of the Sapienza University of Rome, Italy. The channel (7.4 m long, $x$-axis) has a rectangular cross-section 0.35 m high ($z$-axis) and 0.25 m wide ($y$-axis) with lateral walls made of transparent glass to permit optical access (Fig. 1). The development of a neutral boundary layer is promoted by increasing the roughness of the channel bottom via randomly distributed pebbles with average diameter ≈ 5 mm. The water depth and the freestream velocity are 0.16 m and 0.34 m s$^{-1}$, respectively. The obstacle arrays, composed of cubes of height $H$ = 15 mm, are located nearly 5 m downstream of the inlet, where the boundary layer is fully turbulent and nearly independent of the fetch.

### 2.1 Canopy Geometry and Tracer Source Arrangement

Each array of obstacles is designed by means of uniform, sharp-edged cubes glued onto the channel bottom in a staggered pattern (Fig. 1). The test section is located nearly $30H$ downstream of the first cube row, where the internal boundary layer is not significantly dependent on the fetch.

The roughness Reynolds number of the flow, $Re_\tau = u_{*,ref} H / \nu$, ranges from 254 to 285, where $\nu = 10^{-6}$ m$^2$s$^{-1}$ is the kinematic viscosity of water and $u_{*,ref}$ is the reference friction velocity, equal to the average of $\sqrt{-\overline{u'w'}}$ measured in the constant-flux layer (CFL, i.e., where local equilibrium between the momentum flux and velocity gradient holds. It should be noted that this is not the sole way of estimating the friction velocity. For example, some authors (e.g., Castro et al. 2017) estimate the friction velocity by assuming that the measured value of $\overline{u'w'}$ in the region just above the roughness elements is lower than that based on the actual surface stress by a factor around 1.15.). Here, $u$ and $w$ are the streamwise (x-axis) and vertical (z-axis) velocity components, respectively; the prime denotes the fluctuation and the overbar the time average. The values of $Re_\tau$ were then well above the critical value of 70 (e.g., Monin and Yaglom 1971), ensuring that the boundary layer was well within the fully-rough-wall regime.

**Table 1** Geometrical characteristics of the three building arrays and source properties

|  | $\lambda_P = 0.1$ | $\lambda_P = 0.25$ | $\lambda_P = 0.4$ |
|---|---|---|---|
| Element height, $H$ (mm) | 15 | 15 | 15 |
| Distance between the elements, $S$ (mm) | 32 | 15 | 9 |
| Displacement height, $d$ (from Kanda et al. 2013) | $0.46\,H$ | $0.78\,H$ | $0.93\,H$ |
| Reference friction velocity, $u_{*,ref}$ (m s$^{-1}$) | 0.019 | 0.0169 | 0.0173 |
| Source mass rate, $Q$ (kg s$^{-1}$ x10$^{-5}$) | 6.05 | 11.9 | 6.68 |
| Concentration at the source, $c_0$ (kg m$^{-3}$ x10$^{-3}$) | 2.5 | 2.5 | 2.5 |
| Roughness Reynolds number, $Re_\tau$ | 285 | 254 | 260 |

Three obstacle arrangements are investigated corresponding to the plan area indices $\lambda_P = A_P/A_T =$ 0.1, 0.25, and 0.4 (Table 1), where $A_P$ is the plan area occupied by the cubes and $A_T$ is the total lot area. Based on the flow regime classification reported in Grimmond and Oke (1999), the first arrangement refers to the regime of isolated flow ($\lambda_P \lesssim 0.13$), the second one to wake-interference flow ($0.13 \lesssim \lambda_P \lesssim 0.35$), while the third refers to skimming flow ($\lambda_P \gtrsim 0.35$). All the three $\lambda_P$ values fall into the range typically found in real cities (0.1–0.6). Details on the arrangement of the cubical elements for $\lambda_P = 0.25$ are depicted in Fig. 1.

The tracer was released continuously from a pipe inserted through a hole in the centre of the upwind cube belonging to the interrogation area, where the point source was circular (internal diameter 1 mm), $1.67H$ tall and emitted a mixture of Rhodamine-WT and water. The input speed of the mixture at the source is around 30 mm s$^{-1}$, i.e., considerably smaller than the flow speed. Hence, the momentum of the scalar jet at the source could be considered almost negligible if compared to that of the local main flow, ensuring the absence of a significant mechanical rise of the tracer plume. Several preliminary tests were performed by changing the mass rate for each of the three geometrical arrangements and, as expected, the resulting non-dimensional concentration fields were not dependent on the mass rate.

**2.2 Velocity and Concentration Measurements**

The velocity components and the tracer concentration are measured simultaneously on the vertical plane section passing through the longitudinal axis of the channel (green line in Fig. 1). The acquisition facility consists of a green laser (5 W, wavelength 532 nm) emitting a light sheet illuminating the acquisition plane (2 mm thick) and of two synchronized cameras, acquiring 250 frames per second at 1024 x 1280 pixels in resolution. The two cameras are aligned vertically to optimize the framing of the area of interest (102 mm wide (x-axis) and 82 mm high (z-axis)) and to reduce image distortion as much as possible.

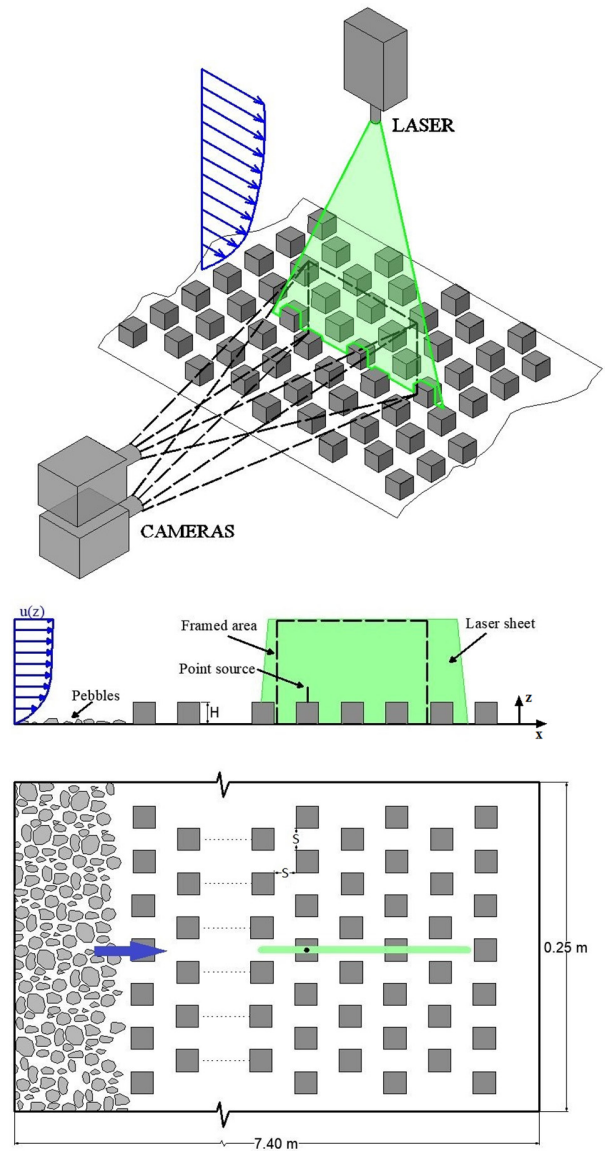

**Fig. 1** Layout of the experimental set-up ($\lambda_P = 0.25$). The green line in the lower panel is the signature along the horizontal plane of the vertical framed area used for the acquisition (central panel)

The first camera acquires the positions of the non-buoyant particles (20 μm in diameter) immersed in the fluid and allows the evaluation of the instantaneous velocity fields by means of a feature-tracking algorithm based on image analysis (Cenedese et al. 2005). A Gaussian interpolation algorithm is applied at each time instant to the scattered samples on the *x–z* plane to obtain the instantaneous velocity field on a regular 204 x 164 array, which corresponds to a 0.5-mm spatial resolution (i.e., *H*/30). To sense only the fluorescent light emission, the second camera is equipped with a narrow band-pass filter tuned on 587 nm (Rhodamine-WT is a passive–chemically inert and neutrally buoyant–fluorescent dye that, excited by a green light (532 nm), emits red light (587 nm)).

The concentration measurements are achieved via planar laser-induced fluorescence (PLIF), a technique typically employed for the investigation of tracer dispersion in water channels. Images are calibrated so that the tracer concentration at a given pixel (directly related to the fractional volume of the dyed fluid) is proportional to the luminosity measured. To allow for the investigation of the scalar flux, the instantaneous concentration field is mapped onto the instantaneous velocity field using an affine transformation accounting for the different points of view of the two cameras. In this process, the superimposition of the velocity and concentration fields is obtained by assigning to each grid cell of the velocity field the value of the concentration at the centre of the cell.

Owing to the flow three-dimensionality, an average of the variables over a sufficiently large number of individual sections belonging to different vertical planes parallel to the streamwise direction would be necessary to obtain representative spatially-averaged properties of the flow. However, for present purposes, we are interested in the mean characteristics of the flow only above the building tops. Hence, it was decided to consider only one vertical section, i.e., the one passing through the centre of the obstacles (green line in Fig. 1). Whilst the spatial averages performed over the latter plane can be considered fully representative of the whole flow when $\lambda_P = 0.25$ (Cheng and Castro 2002), such a simplification might be inappropriate for other geometrical arrangements, although the regular dispositions of the cubes considered in our experiments should not lead to appreciable errors, particularly for $\lambda_P = 0.4$.

## 2.3 Data Analysis

During each experiment, 25,000 instantaneous velocity and concentration fields were measured corresponding to a 100-s long acquisition, i.e., about 200 integral time scales of the turbulence for the present experiments (the Eulerian integral time scale of the streamwise velocity component was measured to be on the order of 0.5 s above the canopy in a different series of measurements conducted in the same configuration by Di Bernardino et al. 2019, hereinafter D19). Therefore, the duration of the experiments was long enough to ensure that the analysis was robust from a statistical point of view. The mean, $\bar{c}$, and the standard deviation, $\sigma_C = \sqrt{\overline{c'^2}}$, of the tracer concentration were determined on the 1024 x 1280 array of the image pixels by applying the canonical Reynolds-averaging method. Similarly, the mean velocity components, $\bar{u}$ and $\bar{w}$, the variances, $\overline{u'^2}$ and $\overline{w'^2}$, as well as the vertical momentum flux, $\overline{u'w'}$, were calculated in each node of the 204 x 164 array. Similarly, the values of the turbulent and the total horizontal (vertical) fluxes of the tracer, $\overline{u'c'}$ and $\overline{uc}$ ($\overline{w'c'}$ and $\overline{wc}$), respectively, were determined on the latter grid.

In what follows, we use dimensionless units, with lengths normalized by the obstacle height, velocities normalized by the friction velocity, and concentrations normalized by a reference value.

## 3 Results and Discussion

### 3.1 Mean and Turbulence Fields

Although the main features of the flow within and above regular arrays of cubical buildings have been well documented (e.g., Cheng and Castro 2002; Kanda et al. 2004; Coceal et al. 2006), for the sake of completeness we report below selected results concerning $\lambda_P = 0.25$, which are useful for the understanding of the dispersive phenomena.

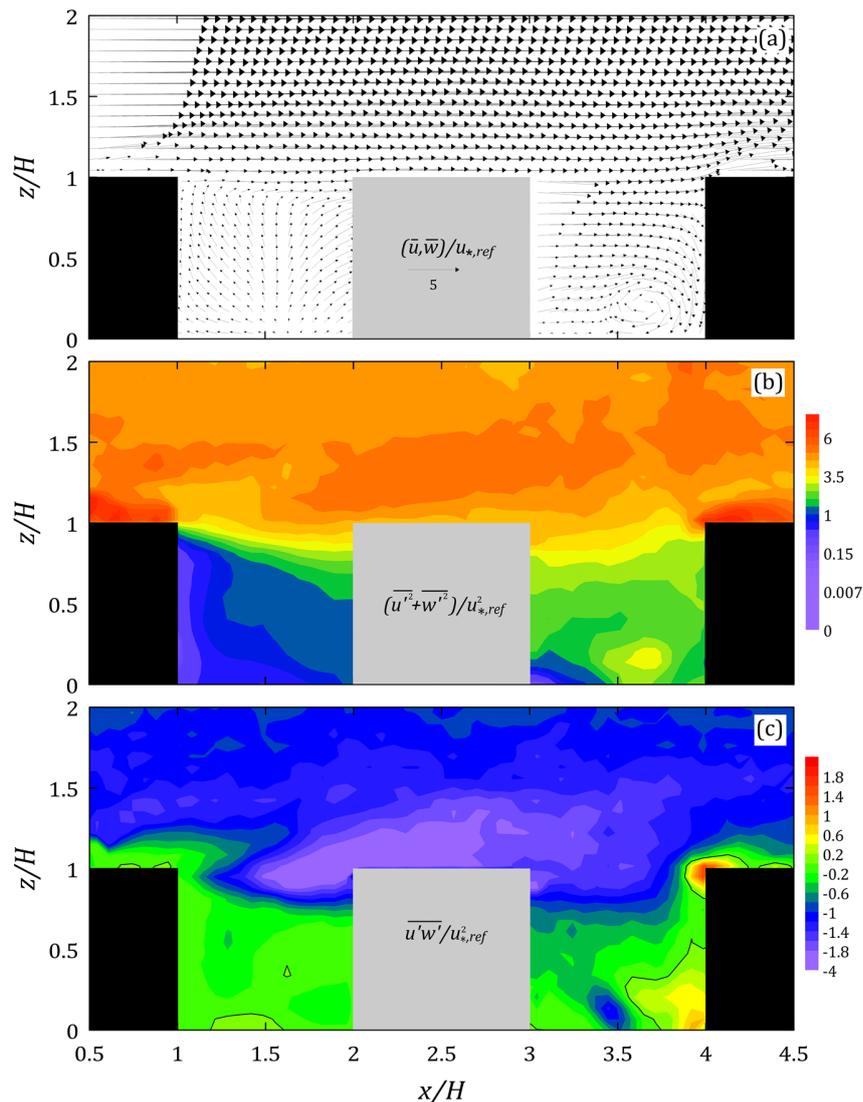

**Fig 2** Normalized (a) mean velocity, (b) velocity variance, and (c) Reynolds stress in the vertical plane passing through the middle of the cubes. The region of optical occlusion due to out-of-plane cubes is shown in grey

Figure 2a shows the vector field of the mean velocity in the vertical plane passing through the middle of the cubes; the values of the two components are normalized by $u_{*,ref}$. The mean flow pattern conforms to the classical configuration already reported in other studies (e.g., Coceal et al. 2006), i.e., a current above the canopy nearly parallel to the streamwise direction, a recirculation in front of the windward wall of the obstacle in the bottom right-hand corner within the canyon, and the absence of a recirculation behind the upwind obstacle.

The map of the non-dimensional flow variance, expressed as $(\overline{u'^2} + \overline{w'^2})/u_{*,ref}^2$, shows that the turbulent activity for $z/H$ is large only near the canopy top and close to the downwind building, in correspondence with the vortical region. The normalized Reynolds stress, $\overline{u'w'}/u_{*,ref}^2$, is negative everywhere except near the top edge and at the right-hand corner of the downwind cube.

Figure 3 illustrates the vertical profiles of the normalized Reynolds shear stress (Fig. 3a) and the standard deviation of the vertical velocity component (Fig. 3b) for the three $\lambda_P$ values. These profiles were estimated by adopting the canopy approach (e.g., Finnigan 2000), i.e., by horizontally averaging the time-averaged statistics over a region including one building top and the contiguous canyon. In doing so, the results can be assumed as representative of the repeating unit constituting the canopy, keeping in mind the limitations previously mentioned regarding cases $\lambda_P = 0.1$ and 0.4. Further results including Eulerian and Lagrangian scales of the turbulence can be found in D19.

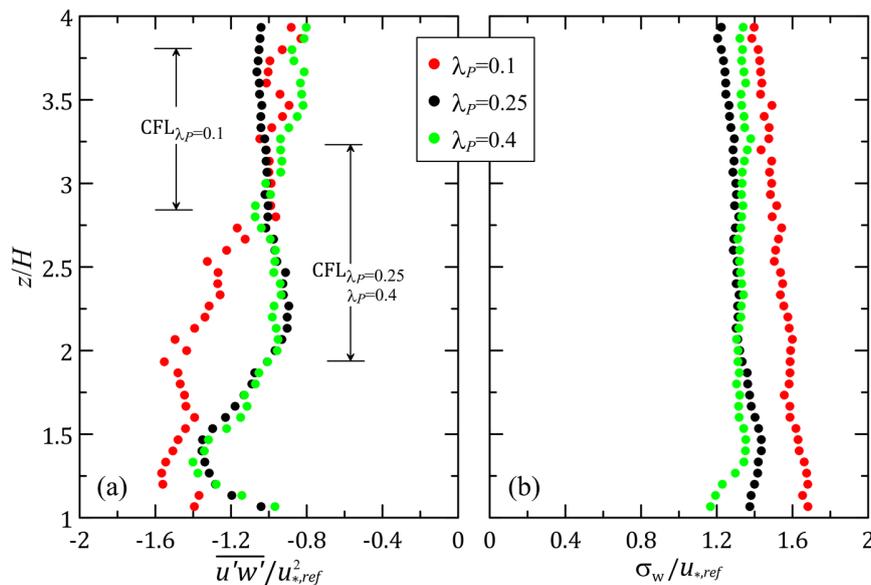

**Fig 3** Vertical profiles of normalized (a) shear stress, and (b) standard deviation of the vertical velocity component for $\lambda_P = 0.1$ (red symbol), $\lambda_P = 0.25$ (black), and $\lambda_P = 0.4$ (green). The reference friction velocities, $u_{*,ref}$, calculated as the averages of the square root of the shear stress in the CFL are listed in Table 1

For $\lambda_P = 0.25$ (black circles), the profiles are quantitatively similar to those reported by other authors (e.g., Cheng and Castro 2002), with $\overline{u'w'}$ (Fig. 3a) varying up to $z \approx 2.75H$ for $\lambda_P = 0.1$ and up to $z \approx 2H$ for $\lambda_P = 0.25$ and 0.4, i.e., the roughness sublayer (RSL) top. The latter is the region immediately above the urban canopy, where the flow is non-homogenous and strongly influenced by the elements that constitute the urban canopy (Rotach 1999). Above the RSL $\overline{u'w'}$ is independent of the height up to $z \approx 3.2H$, which can be considered as the CFL top.

Whilst no significant difference is apparent between $\lambda_P = 0.4$ (skimming flow) and 0.25, for $\lambda_P = 0.1$ (isolated regime) the RSL is considerably deeper and the CFL forms at $z \approx 2.8H$. This agrees with other observations conducted both in the laboratory and in the real field (Salizzoni et al. 2011; Pelliccioni et al. 2016). Note also that $\sigma_w/u_{*,ref}$ does not change appreciably with height in the whole $z/H$ range analyzed here, with the $\lambda_P = 0.1$ case showing slightly larger values, in agreement with Leonardi and Castro (2010).

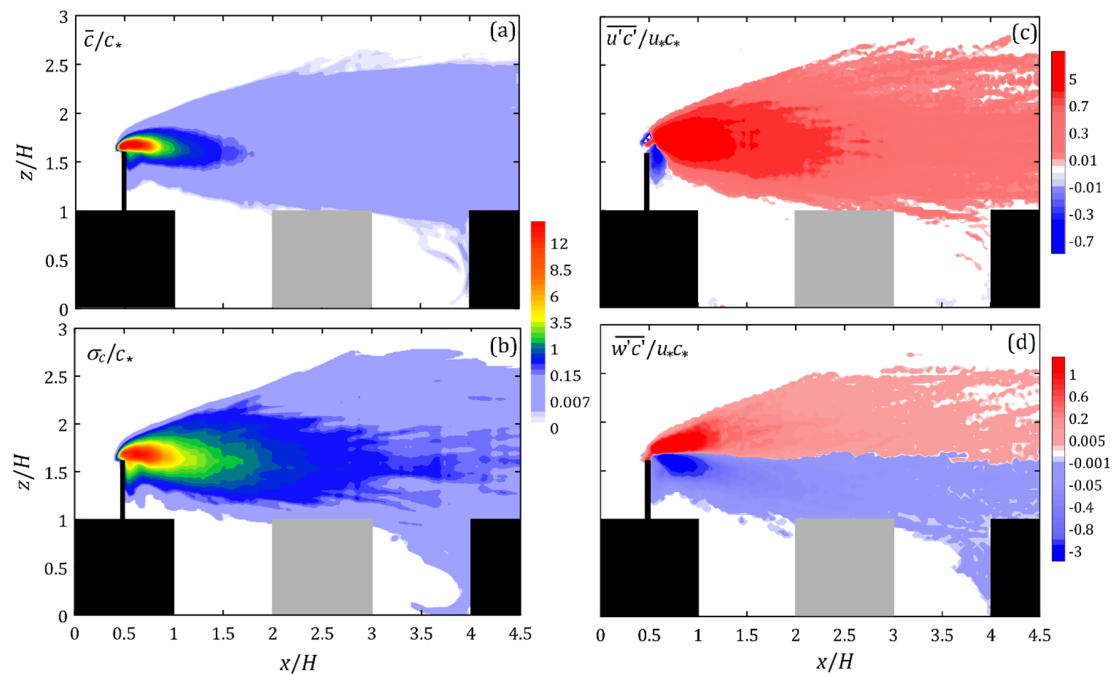

**Fig 4** Panels (a) and (b) show the maps of the mean and standard deviation of the normalized concentration statistics, while maps (c) and (d) depict the normalized turbulent tracer flux along the horizontal and vertical directions. Please note that, for the sake of readability, the maps of panels (c) and (d) use different colour scales

### 3.2 Mean and Standard Deviation of Concentration

The tracer concentration is normalized by the reference concentration, $c_* = Q/(H^2 u_{*,ref})$, where $Q$ is the mass rate of the source, which is located within the RSL for all the three obstacle arrangements ($z = 1.67H$).

As expected, the plume centreline is nearly parallel to the streamwise direction and only a small portion of it enters into the canopy, essentially by means of flow advection at the windward wall of the downwind building (Fig. 4a). The shape of the plume associated with the normalized standard deviation of concentration ($\sigma_C/c_*$, Fig. 4b) is similar to that of the mean. At the source, $\sigma_C/c_*$ is zero because there is no concentration fluctuation there. In the proximity of the source, the deformation of the lower boundary of the plume induced by the wake of the stack is apparent, carriyng pollutants downward and produces a slight lowering of the plume axis. The largest $\sigma_C/c_*$ are found near the plume axis and are of the same order of $\bar{c}/c_*$.

### 3.3 Turbulent Mass Fluxes

The expected values of the horizontal (vertical) tracer flux $\overline{uc}$ ($\overline{wc}$) is expressed as the sum of the mean (or advective) flux, $\bar{u}\bar{c}$ ($\bar{w}\bar{c}$), and the turbulent (or eddy) flux, $\overline{u'c'}$ ($\overline{w'c'}$). Figures 4c and 4d depict $\overline{u'c'}$ and $\overline{w'c'}$ normalized by $u_*c_*$, respectively, while some of the vertical profiles of the same variables taken at several distances downwind of the source are shown in Figs. 5 and 6.

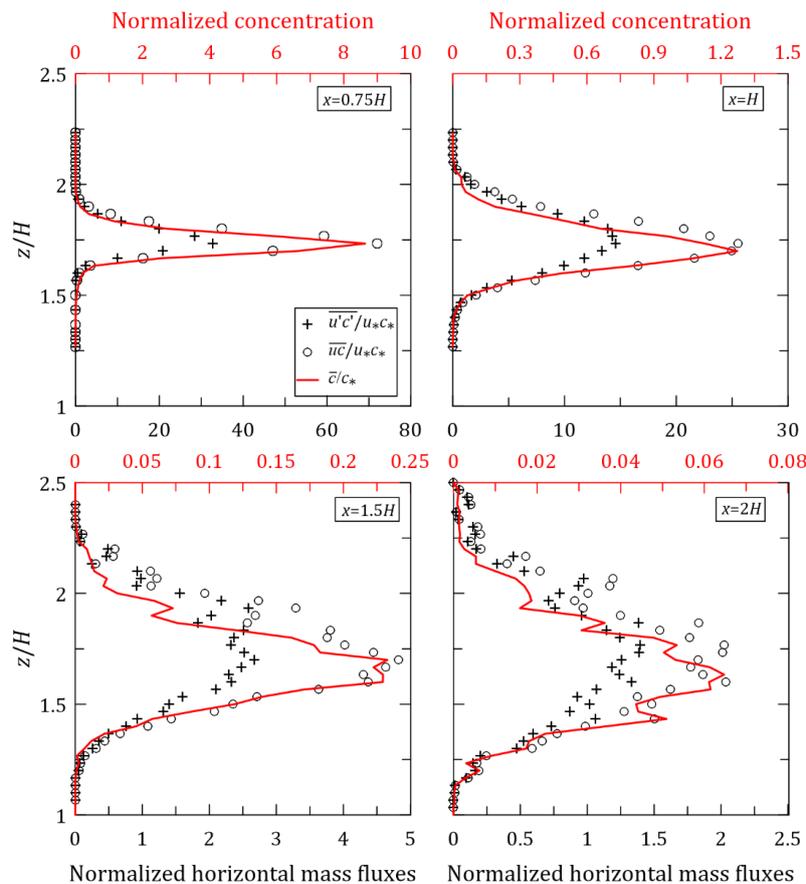

**Fig 5** Vertical profiles of the normalized concentrations ($\bar{c}/c_*$, red lines), horizontal turbulent ($\overline{u'c'}/u_*c_*$, crosses), and total ($\overline{uc}/u_*c_*$, circles) mass fluxes measured at four downwind distances from the tracer source

Except within the wake of the stack, $\overline{u'c'}/u_*c_*$ is positive everywhere (Fig. 4a), with the largest values observed at the plume axis. Since $\partial \bar{c}/\partial x < 0$ within the plume (not shown), $\overline{u'c'}/u_*c_*$ is everywhere downgradient. Note that the ratio $\overline{u'c'}/\overline{uc} \approx 0.5$ for the area close to the plume axis, i.e., for $z/H \approx 1.75$ (Fig. 5); with regard to $\overline{w'c'}$, it is positive (negative) above (below) the plume axis and generally lower than $\overline{u'c'}$ (Fig. 4d). In contrast, given the small mean vertical velocity, $\overline{wc} \approx \overline{w'c'}$ except in the region close to the source ($x = 0.75H$ and $x = H$), where the plume downwash produces a slight asymmetry in the pattern of the plume of $\overline{wc}$ (Fig. 6). The most obvious feature in Fig. 6 is the correspondence of the location of the maximum $\bar{c}$ value with that of the change in sign of $\overline{w'c'}$. This suggests that also the vertical turbulent flux is always downgradient and thus positive $\overline{w'c'}$ values are associated with negative $d\bar{c}/dz$ (and vice versa), consistent with the analogy of molecular behaviour.

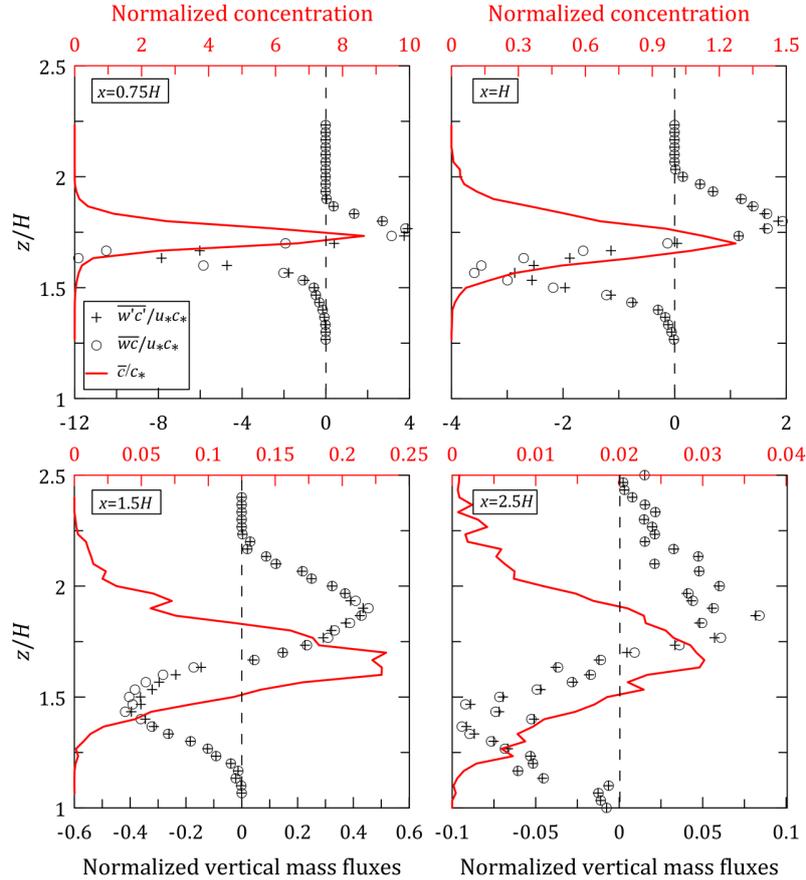

**Fig 6** As in Fig. 5, but for the vertical component of the tracer fluxes

### 3.4 The Turbulent Schmidt Number

The turbulent Schmidt number is defined as the ratio of the eddy diffusivity of momentum to the eddy diffusivity of mass, i.e.,

$$Sc_t = \frac{K_M}{D_t}. \tag{1}$$

The diffusivities $K_M$ and $D_t$ can be formulated using gradient transport theory (or $K$-theory), viz.

$$K_M = -\frac{\overline{u'w'}}{\partial \bar{u}/\partial z} \tag{2a}$$

$$D_t = -\frac{\overline{w'c'}}{\partial \bar{c}/\partial z}. \tag{2b}$$

Given the hypothesis of horizontal homogeneity of the flow field, a vertical profile of $K_M$ can be readily calculated from the horizontally-averaged vertical profiles of $\overline{u'w'}$ and $\bar{u}$ shown in Fig. 3. The obvious question then arises as to the required locations of the vertical profiles of $\overline{w'c'}$ and $\bar{c}$ must have for the computation of $D_t$. Indeed, since the width of the polluted plume close to the source is generally lower than the length scale of the turbulence structures characterizing the flow, $D_t$ invariably also depends on the streamwise position (for a comprehensive discussion on this issue, see Fackrell and Robins 1982; Harman and Finnigan 2008). It is also worthwhile recalling that *K*-theory fails when larger-size eddies are present in the flow (Stull 1988). However, the absence of counter-gradient fluxes in our experiments leads us to assume that the *K*-theory approximates $\overline{w'c'}$ in a reasonable way, although it is physically unrealistic in principle. With this in mind, we calculated $D_t$ using Eq. 2 and the vertical profiles of $\bar{c}$ and $\overline{w'c'}$ of the kind reported in Fig. 6 passing through $x/H$ = 2.5. This can be considered a sound compromise between the reliability of concentration data and plume width, the latter about four times the vertical length scale of the turbulence in the present flow regimes (see D19).

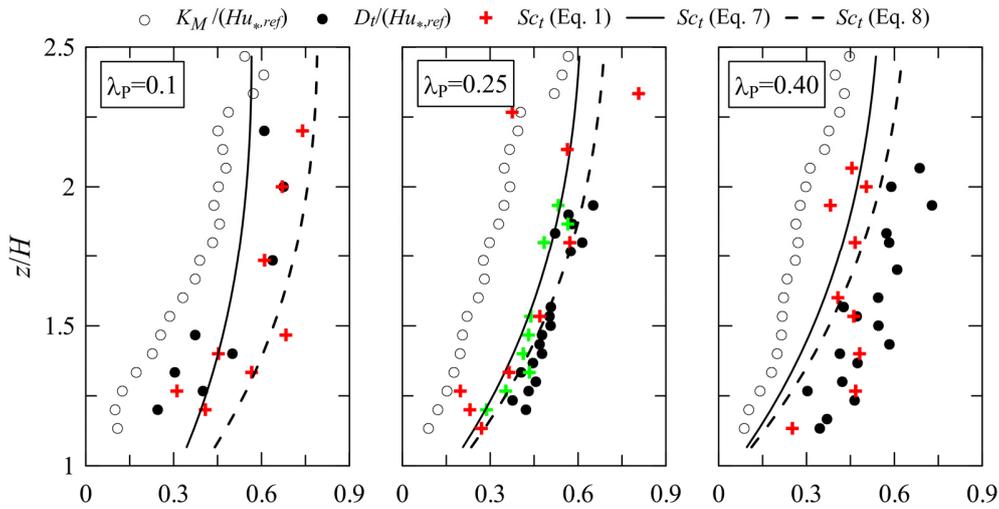

**Fig 7** Vertical profiles of the normalized eddy diffusivity of momentum (horizontal average, open circles), eddy diffusivity of mass (calculated at $x/H$ = 2.5, solid circles), and the corresponding turbulent Schmidt numbers estimated by Eq. 1 (red crosses), by Eq. 7 (continuous line), and by Eq. 8 (dashed line) for the three $\lambda_P$. The green crosses in the central panel refer to $Sc_t$ calculated by Eq. 1 using the eddy diffusivity of mass at $x/H$ = 1.5

The vertical profiles of $K_M$ (open circles, from D19), $D_t$ (solid circles), and $Sc_t$ (red crosses) for the three canopies are shown in Fig. 7. It should be noted that the computation of the eddy mass flux and the spatial derivatives of the mean concentration may be affected by significant errors at the borders of the plume, where the concentrations are very low. Therefore, we excluded such regions from our discussion. Furthermore, to reduce errors due to unavoidable oscillations of $\bar{c}$ along the vertical profiles, which have unwanted effects on the estimations of $\partial \bar{c}/\partial z$, we interpolated $\bar{c}$ using a moving average over a distance equal to $H/15$. Finally, unrealistic $D_t$ values due to $\partial \bar{c}/\partial z \approx 0$ around the $\bar{c}$ maximum have been removed.

The results show a clear increase of $Sc_t$ with height regardless of $\lambda_P$. This agrees with Koeltzsch (2000), who measured $Sc_t$ in a wind tunnel in the case of flat terrain, finding an $Sc_t$ maximum around 0.9 at one-third of the boundary-layer height and $Sc_t \approx 0.3$ near the surface.

Despite the fact that the data scatter makes the interpretation of our results uncertain at the upper edge of the plume, larger $Sc_t$ ($\approx 0.8$) values occur for $\lambda_P = 0.1$, while smaller values ($\approx 0.6$) take place for $\lambda_P = 0.25$ and 0.4. On the other hand, $Sc_t \approx 0.3$ nearby the canopy top. It is worth mentioning that $Sc_t \approx 0.6$ is not far from the (constant) $Sc_t$ found in the real field by Flesh (2000) and Wilson (2013) or from $Sc_t = 0.6 - 0.7$ based on Project Prairie Grass measurements (see e.g., Wilson and Yee 2007). For a comprehensive discussion on the $Sc_t$ variability in the RSL, see also Townsend (1979), Rogers (1991), and Wilson (2013). Lastly, Fig. 7 also shows the $Sc_t$ profile calculated at $x = 1.5H$ for $\lambda_P = 0.25$ (green crosses in the central panel). As is evident, it is substantially coincident with values obtained at $x = 2.5H$ (red crosses), thus reinforcing the hypothesis that the estimated $Sc_t$ does not depend appreciably on the streamwise direction for the downstream distances we analyzed in the laboratory.

**3.5 Turbulent Schmidt Number Modelling**

With a view to finding a simple formulation for $Sc_t$ as an alternative to Eq. 1, which also could be used in numerical models, we followed an approach similar to that of Flesh et al. (2002). In particular, we modelled the eddy diffusivity of momentum over the canopy by appealing to Prandtl's mixing-length theory, $K_M = ku_{*,ref}(z - d)$, the latter to be assumed valid, in principle, only within the CFL (here $k = 0.41$ is the von Kármán constant and $d$ the displacement height). Instead of the *K*-theory, we use for $D_t$ the expression for the far-field eddy diffusivity of mass derived by Sawford and Guest (1988) based on the criteria given by Thomson (1987) in the framework of Lagrangian stochastic model applications of particle trajectories in turbulent flows. In particular, $D_t = 2(u_{*,ref}^4 + \sigma_{w,ref}^4)/(C_0\varepsilon)$, where $\sigma_{w,ref}$ denotes $\sigma_w$ calculated in the CFL and $C_0$ is the Kolmogorov constant, whose value in inhomogeneous and anisotropic turbulence is generally assumed to fall within the range 2–7 (see Poggi et al. 2008 for a review on $C_0$ values found in the literature). Therefore, Eq. 1 now reads

$$Sc_t = \frac{ku_{*,ref}(z - d)}{2(u_{*,ref}^4 + \sigma_{w,ref}^4)} C_0\varepsilon, \qquad (3)$$

noting that knowledge of $C_0\varepsilon$ is also of great interest in Lagrangian dispersion model applications in that it appears in the stochastic term of the generalized Langevin equation (see e.g., Anfossi et al. 2006; Amicarelli et al. 2011). For that reason too, we calculate $C_0\varepsilon$ by using the Tennekes (1982) relationship

$$C_0 \varepsilon = \frac{2\sigma^2}{T^L}, \qquad (4)$$

where $T^L$ is the Lagrangian time scale of turbulence. Figure 8 shows the vertical profiles of $C_0\varepsilon$ normalized by $u_{*,ref}^3/H$ calculated using Eq. 4 and $T^L$ estimated by D19 for the three $\lambda_P$ (circles). It is interesting to observe that the three profiles collapse onto a single curve as if $C_0\varepsilon$ follows a sort of self-similarity. In all three experiments, $C_0\varepsilon H/u_{*,ref}^3$ is well approximated by an exponential relation of the form

$$\frac{C_0 \varepsilon H}{u_{*,ref}^3} = \alpha \exp\left[-\beta \frac{z}{H}\right], \qquad (5)$$

which corresponds to the line in Fig. 8 ($\alpha$ = 22, $\beta$ = 0.5, $R^2 = 0.97$).

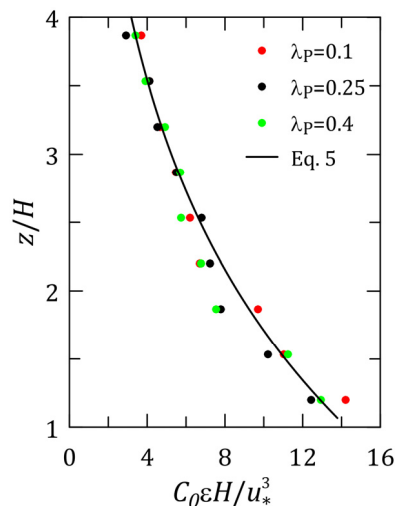

**Fig 8** Vertical profiles of $C_0\varepsilon H/u_{*,ref}^3$ for the three $\lambda_P$. The line depicts Eq. 5

Substituting Eq. 5 in Eq. 3, one obtains an expression for $Sc_t$ as a function of height

$$Sc_t = \frac{k u_{*,ref}^4}{2(u_{*,ref}^4 + \sigma_{w,ref}^4)} \left(\frac{z-d}{H}\right) \alpha \exp\left[-\beta \frac{z}{H}\right]. \qquad (6)$$

Since $\sigma_{w,ref} \approx A u_{*,ref}$, where $A$ is a constant of order 1 (Panofsky and Dutton 1984), a simpler expression for $Sc_t$ can be readily obtained from Eq. 6

$$Sc_t = \frac{k\alpha}{2(1 + A^4)} \left(\frac{z-d}{H}\right) \exp\left[-\beta \frac{z}{H}\right], \qquad (7)$$

which gives the vertical profile of the turbulent Schmidt number above the obstacles from known geometrical characteristics of the canopy ($d$ and $H$) and the parameter $A$. The continuous lines in Fig. 7 depict $Sc_t$ from Eq. 7 for the three $\lambda_P$ values. Its agreement with Eq. 1 (red crosses) and Eq. 7 is reasonably good for $\lambda_P = 0.25$ ($A \approx 1.3$), while a certain discrepancy occurs for $z < 1.5H$ for $\lambda_P = 0.4$ ($A \approx 1.3$) and $z > 1.75H$ for $\lambda_P = 0.1$ ($A \approx 1.45$).

A switch from the effective $A$ to its canonical value 1.2 generally assumed in the CFL (Stull 1988) yields

$$Sc_t = 1.4 \left(\frac{z-d}{H}\right) \exp\left[-\beta \frac{z}{H}\right]. \tag{8}$$

Such a simpler and easy-to-use form of Eq. 7 requires no information on the velocity field. The agreement between Eq. 8 (dashed line) and the experiments remains satisfactory for all three obstacle arrangements.

## 4 Comparisons with Other $Sc_t$ Formulations

Whilst most previous studies on pollutant dispersion in canopy flows have focused on the optimum (constant) $Sc_t$ value to best calculate the concentration field in common RANS models, less attention has been paid to the definition of relationships relating $Sc_t$ and the flow field. As mentioned in the introduction, one interesting attempt was made by Gorlé et al. (2010), who proposed an expression for $Sc_t$ valid in homogeneous, stationary, and isotropic turbulence, viz.

$$Sc_t = \frac{9}{8} C_\mu C_0, \tag{9}$$

where $C_\mu$ is the constant of proportionality between the eddy diffusivity and $q^2/\varepsilon$ (see Sect. 1) and

$$C_0 = \frac{C_{0\infty}}{1 + 7.5 C_{0\infty}^2 Re_\lambda^{-1.64}}, \tag{10}$$

where $Re_\lambda = \lambda \sigma_u / \nu$ is the Taylor microscale Reynolds number, $\sigma_u$ is the standard deviation of the streamwise velocity component, and $\lambda$ is the Taylor microscale. Here, Gorlé et al. (2010) denoted $C_{0\infty}$ as the Kolmogorov constant (to be considered independent of the Reynolds number and assumed equal to 6 by those authors). However, since for practical applications $Re_\lambda$ is large ($\gtrsim 100$), $C_0$ does not vary much throughout the flow field. As a consequence, the resulting value for $Sc_t$ found by Gorlé et al. (2010) in their case study (pollutant dispersion in the wake of a single rectangular building) ranged in the 0.42–0.54 interval.

More recently, Longo et al. (2019) proposed an alternative formulation for $Sc_t$, viz.

$$Sc_t = \frac{2C_\mu}{C_0 C^2}. \tag{11}$$

These authors used parameters similar to those that appear in the Gorlé et al. model, but assumed $C_\mu$ dependent both on the distance from the wall and on the strain-rate tensor and rotation matrix (Ehrhard and Moussiopoulos 2000), accounting for the position within the flow and the local turbulence intensity. Furthermore, in the Longo et al. model the coefficients $C_0$ and $C$ were determined based on the exploration of the uncertainty range of those parameters, its effect on $Sc_t$ and concentration field, and the definition of a least square point with respect to available experimental data. For details on the procedure, refer to Longo et al. (2019). The optimal values they found for the parameters were $C_0 = 2$ and $C = 0.35$. Note that both the previous authors used the commercial CFD solver ANSYS Fluent for their analysis.

Figure 9 shows the vertical profiles of $Sc_t$ calculated using the Gorlé et al. model (Eqs. 9 and 10), the Longo et al. model (Eq. 11) and our formulation (Eq. 8). Since in our experiments we did not measure the standard deviation of the velocity component along the spanwise direction, we could not use the expression for the (constant) $C_\mu$ based on $q$ as done by Gorlé et al. (2010). Hence, we set $C_\mu = 0.09$ as typically assumed in ANSYS Fluent, while the Taylor microscale was calculated as $\lambda = \sqrt{15\sigma_u^2 \nu/\varepsilon}$ (see Di Bernardino et al. 2017 for details).

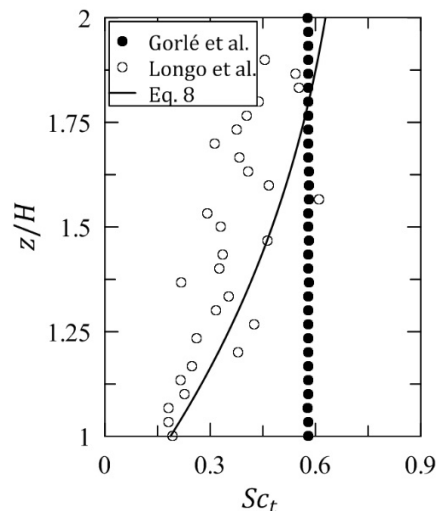

**Fig 9** Vertical $Sc_t$ profiles calculated by Eq. 8 (continuous line), Gorlé et al.'s (2010) model (solid circles), and Longo et al.'s (2019) model (open circles) for $\lambda_P = 0.25$

As expected, for the Gorlé et al. model $Sc_t$ is nearly constant (solid circles), showing a value of about 0.6 over the whole height range considered, not far from our experimental values at the upper $z$-levels. On the other hand, the Longo et al. model shows a clear growth with the height (open circles), consistent with Eq. 8 (line). Despite the inevitable data scatter caused by oscillations in $C_\mu$, there is reasonable agreement between the laboratory and the Longo et al. $Sc_t$ model, particularly for $z < 1.5H$.

## 5 Concluding Remarks

The main conclusions that can be drawn from our water-channel experiments are:

i) The growth with height we found for $Sc_t$ is further evidence that the turbulent Schmidt number cannot be considered as a constant above the canopy top, particularly within the roughness sublayer. To the authors' knowledge, so far there have been no $Sc_t$ investigations conducted experimentally above staggered arrays of cubic obstacles.
ii) The normalized form of the factor $C_0\varepsilon$ (i.e., the product of the Kolmogorov constant by the rate of dissipation of the turbulence kinetic energy) decreases exponentially with height and, interestingly, it is substantially independent of the plan-area fraction of the array.
iii) We found that $Sc_t$ and the height are related by Eq. 7, which does not need any information on $C_0$, $\varepsilon$, or the integral time scales of the turbulence, in contrast with previous expressions proposed in the literature. This is a significant point of strength since, as is widely known, those parameters are difficult to evaluate, particularly in the real field.
iv) The simpler formulation of Eq. 8, which is based only on geometrical parameters, performs reasonably well for all the three cube arrays investigated in this work, at least for the downstream distances we analysed. This result could be useful to numerical modellers when setting the turbulent Schmidt number above urban canopies.

**Acknowledgments** The assistance of Manuel Mastrangelo and Cristina Grossi (Master degree students of the University of Rome "La Sapienza") to the measurements was greatly appreciated. This research was supported by the RG11715C7D43B2B6 fund from the University of Rome "La Sapienza".